\let\ifarxiv=\iftrue     
\pdfoutput=1

\ifarxiv

\documentclass[12pt,a4paper]{article}
\usepackage[a4paper,text={450pt,650pt},centering]{geometry}

\fi

\ifarxiv\else

\documentclass[11pt,a4paper]{article}
\usepackage{mathptmx}
\usepackage[a4paper,text={130mm,198mm}]{geometry}

\fi

\usepackage{amsmath,amssymb}
\usepackage[bookmarks=true,hyperfigures=true]{hyperref}
\usepackage{graphicx}
\usepackage[nosort]{cite}
\usepackage[bulletsep]{collref}

\let\oldbfseries=\bfseries
\let\oldmdseries=\mdseries
\let\oldnormalfont=\normalfont
\renewcommand{\bfseries}{\oldbfseries\boldmath}
\renewcommand{\mdseries}{\oldmdseries\unboldmath}
\renewcommand{\normalfont}{\oldnormalfont\unboldmath}

\allowdisplaybreaks[3]

\numberwithin{equation}{section}

\usepackage[font=small,labelfont=bf,width=0.85\textwidth]{caption}

\providecommand{\hypersetup}[1]{}

\hypersetup{plainpages=false}
\hypersetup{pdfpagemode=UseNone}
\hypersetup{bookmarksnumbered=true}
\hypersetup{pdfstartview=FitH}
\hypersetup{colorlinks=false}
\hypersetup{citebordercolor={.5 1 .5}}
\hypersetup{urlbordercolor={.5 1 1}}
\hypersetup{linkbordercolor={1 .7 .7}}

\providecommand{\href}[2]{#2}
\providecommand{\arxivlink}[1]{\href{http://arxiv.org/abs/#1}{arxiv:#1}}

\def\AA{\mathcal{A}}
\def\BB{\mathcal{B}}
\def\CC{\mathcal{C}}
\def\cZ{\mathcal{Z}}
\def\cW{\mathcal{W}}

\begin{document}


\thispagestyle{empty}
\phantomsection
\addcontentsline{toc}{section}{Title}

\begin{flushright}\footnotesize%
\texttt{CERN-PH-TH/2010-307},
\texttt{\arxivlink{1012.4002}}\\
overview article: \texttt{\arxivlink{1012.3982}}%
\vspace{1em}%
\end{flushright}

\begingroup\parindent0pt
\begingroup\bfseries\ifarxiv\Large\else\LARGE\fi
\hypersetup{pdftitle={Review of AdS/CFT Integrability, Chapter V.2: Dual Superconformal Symmetry}}%
Review of AdS/CFT Integrability, Chapter V.2:\\
Dual Superconformal Symmetry
\par\endgroup
\vspace{1.5em}
\begingroup\ifarxiv\scshape\else\large\fi%
\hypersetup{pdfauthor={J. M. Drummond}}%
J.~M.~Drummond
\par\endgroup
\vspace{1em}
\begingroup\itshape
PH-TH Division, CERN, 1211 Geneva 23, Switzerland\par\vspace{1em}
LAPTH, Universit\'{e} de Savoie, CNRS,
B.P.\ 110,  F-74941 Annecy-le-Vieux Cedex, France
\par\endgroup
\vspace{1em}
\begingroup\ttfamily
drummond@lapp.in2p3.fr
\par\endgroup
\vspace{1.0em}
\endgroup

\begin{center}
\includegraphics[width=5cm]{TitleV2.mps}
\vspace{1.0em}
\end{center}

\paragraph{Abstract:}
Scattering amplitudes in planar $\mathcal{N}=4$ super Yang-Mills theory reveal a remarkable symmetry structure. In addition to the superconformal symmetry of the Lagrangian of the theory, the planar amplitudes exhibit a dual superconformal symmetry. The presence of this additional symmetry imposes strong restrictions on the amplitudes and is connected to a duality relating scattering amplitudes to Wilson loops defined on polygonal light-like contours. The combination of the superconformal and dual superconformal symmetries gives rise to a Yangian, an algebraic structure which is known to be related to the appearance of integrability in other regimes of the theory. We discuss two dual formulations of the symmetry and address the classification of its invariants.

\ifarxiv\else
\fi
\hypersetup{pdfsubject={MSC (2010): 81R25, 81R50, 81T13, 81T60}}%

\ifarxiv\else
\fi
\hypersetup{pdfkeywords={Gauge theories, Scattering amplitudes, Wilson loops, Extended symmetries.}}%

\newpage




\section{Introduction}

The aim of this article is to give an overview of the role of extended symmetries in the context of scattering amplitudes in $\mathcal{N}=4$ super Yang-Mills. We will begin by examining the structure of the loop corrections in perturbation theory. The scattering amplitudes are typically described in terms of scalar loop integrals. The integrals contributing in the planar limit turn out to reveal a remarkable property, namely that when exchanged for their dual graphs they exhibit a new conformal symmetry, dual conformal symmetry. 

This symmetry of scattering amplitudes is also revealed if one considers the strong coupling description which is given in terms of minimal surfaces in AdS${}_5$. More detail on this subject can be found in \cite{chapTdual}.
A T-duality transformation of the classical string equations of motion then relates scattering amplitudes to Wilson loops defined on polygonal light-like contour. The T-dual space where the Wilson loop is defined is related to the momenta of the particles in the scattering amplitude. 
The relation to Wilson loops has been observed for certain amplitudes also in the perturbative regime. From the symmetry point of view the most important consequence of this is that the conformal symmetry naturally associated to the Wilson loop also acts as a new symmetry of the amplitudes. Thus the dual conformal symmetry is at least partially explained by the duality between amplitudes and Wilson loops. The explanation is by no means complete as the dual description only applies to the special class of maximally-helicity-violating (MHV) amplitudes. However it turns out that the notion of dual conformal symmetry seems to apply to all amplitudes and furthermore naturally extends to a full dual superconformal symmetry. In particular tree-level amplitudes of all helicity types are covariant under dual superconformal symmetry. We will describe the formulation of the these symmetries and discuss to what extent the symmetry is controlled beyond tree-level.

The combination of the original Lagrangian superconformal symmetry and the dual superconformal symmetry yields a Yangian structure. This structure arises in many regimes of the planar AdS/CFT system and can be thought of as the indicator of the integrability of the model. A natural question which arises with such an infinite-dimensional symmetry to hand is whether one can classify all of its invariants. It turns out that a remarkable integral formula which gives all possible leading singularities of the perturbative scattering amplitudes also fills the role of providing all possible Yangian invariants. In some sense this indicates that the planar amplitude is being determined by its symmetry at the level of its leading singularities. More concretely one can say that the integrand of the amplitude is Yangian invariant up to a total derivative. It remains to be seen to what extent these ideas can be extended to determine the loop corrections themselves, i.e. after doing the loop integrations.

\section{Amplitudes at weak coupling}

We will begin our discussion by examining perturbative scattering amplitudes in $\mathcal{N}=4$ super Yang-Mills theory in the planar limit. Further details on scattering amplitudes in perturbation theory can be found in \cite{chapAmp}.
A lot can be learned from the simplest case of the four-gluon scattering amplitudes. Due to supersymmetry, the only non-zero amplitudes are those for two gluons of each helicity type. These amplitudes are examples of the so-called maximally-helicity-violating or MHV amplitudes which have a total helicity of $(n-4)$. For four particles this quantity vanishes and so by applying a parity transformation one can see that the amplitudes are also anti-MHV or $\overline{\rm MHV}$ amplitudes. MHV amplitudes are particularly simple in that they can naturally be written as a product of the rational tree-level amplitude and a loop-correction function which is a series in the `t Hooft coupling $a$,
\begin{equation}
\mathcal{A}_n^{\rm MHV} = \mathcal{A}_{n,{\rm tree}}^{\rm MHV} \, M_n(p_1,\ldots,p_n;a).
\end{equation}
One can write any amplitude in this form of course, but the special property of MHV amplitudes is that the function $M_n$ is a function which produces a constant after taking $2l$ successive discontinuities at $l$ loops. In other words, all leading singularities of MHV amplitudes are proportional to the MHV tree-level amplitude. Strictly speaking the amplitude is infrared divergent and the function $M_n$ also depends on the regularisation parameters. The operation of taking $2l$ discontinuities at $l$ loops yields an infrared finite quantity however and the regulator can therefore then be set to zero. 

\begin{figure}[htbp]
\centerline{\includegraphics[height=5cm]{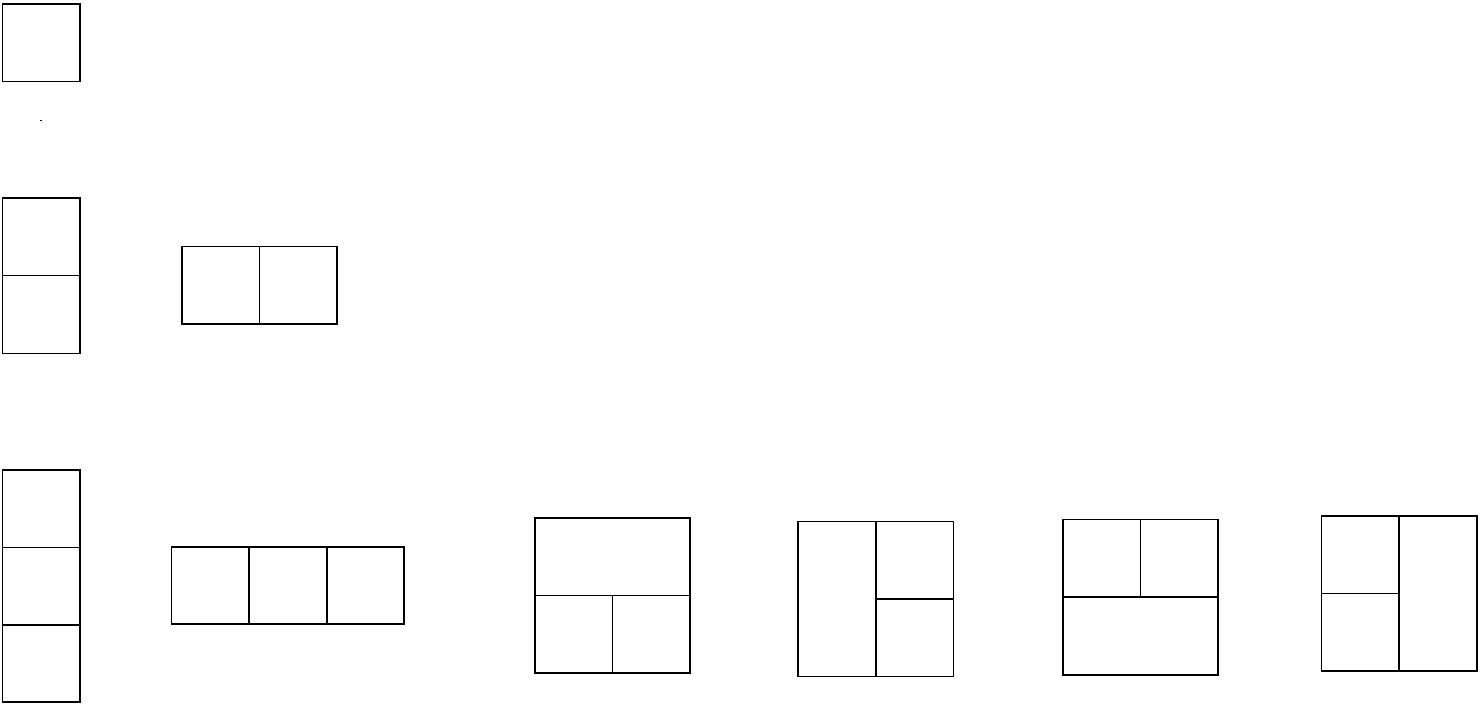}}
\caption{Integral topologies up to three loops. The external momenta flow in at the four corners in each topology.}
\label{figure:4ptamp}
\end{figure}

\begin{figure}[htbp]
\ifarxiv
\centerline{\includegraphics[height=5cm]{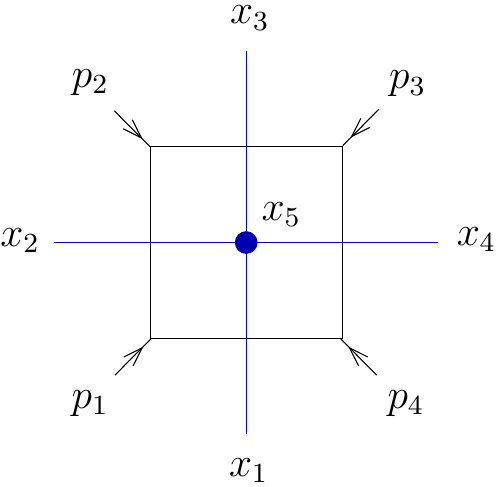}} 
\else
\centerline{\includegraphics[height=5cm]{dualdiagtimes}} 
\fi
\caption{Dual diagram
for the one-loop box. The black lines denote the original momentum space loop integral. The propagators can equivalently be represented as scalar propagators in the dual space, denoted by the blue lines.}
\label{figure:dualdiag}
\end{figure}

\begin{figure}[htbp]
\centerline{\includegraphics[height=4cm]{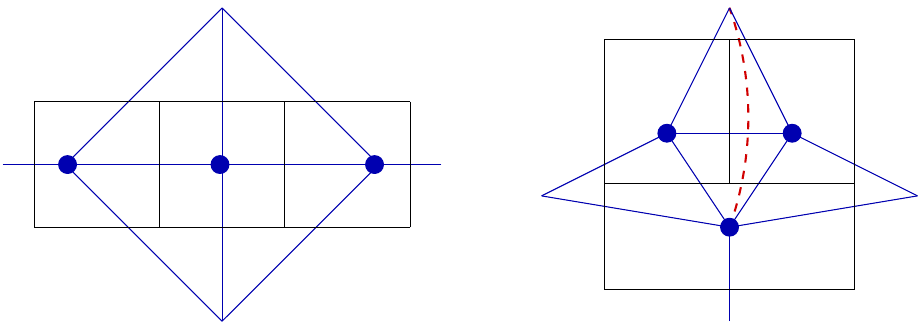}}
\caption{Dual diagrams for the three-loop box and for the `tennis court'
with its numerator denoted by the dashed line corresponding to a factor in the numerator of the square distance between the two points.} \label{figure:3ladandtc}
\end{figure}

The function $M_n$ is given by a perturbative expansion in terms of scalar loop integrals. If we consider the four-particle case then the relevant planar loop integral topologies appearing up to three-loop order are of the form shown in Fig. \ref{figure:4ptamp} \cite{Anastasiou:2003kj,Bern:2005iz}
The integrals contributing to $M_4$ all have a remarkable property - they exhibit an unexpected conformal symmetry called `dual conformal symmetry' \cite{Drummond:2006rz}. The way to make this symmetry obvious is to make a change of variables from momentum parametrisation of such integrals to a dual coordinate representation,
\begin{equation}
p_i^\mu = x_i^\mu - x_{i+1}^\mu \equiv x_{i,i+1}^\mu, \qquad x_{n+1}^\mu \equiv x_1^{\mu}.
\label{dualcoords}
\end{equation}
We will illustrate this here on the example of the one-loop scalar box integral which is the one-loop contribution to $M_4$,
\begin{equation}\label{1box}
    I^{(1)} = \int \frac{d^4k}{k^2(k-p_1)^2(k-p_1-p_2)^2(k+p_4)^2}\ .
\end{equation}
In this case the change of variables takes the form,
\begin{equation}
p_1 = x_{12}, \qquad p_2 = x_{23}, \qquad p_3 = x_{34}, \qquad p_4 = x_{41}.
\end{equation}
The integral can then be written as a four-point star diagram (the dual graph for the one-loop box) with the loop integration replaced by an integration over the internal vertex $x_5$ as illustrated in Fig. \ref{figure:dualdiag}.
In the new variables a new symmetry is manifest. If we consider conformal inversions of the dual coordinates,
\begin{equation}
x_i^\mu \longrightarrow - \frac{x_i^{\mu}}{x_i^2},
\end{equation}
then we see that the integrand, including the measure factor $d^4x_5$, is actually covariant,
\begin{equation}
\frac{d^4 x_5}{x_{15}^2 x_{25}^2 x_{35}^2 x_{45}^2} \longrightarrow (x_1^2 x_2^2 x_3^2 x_4^2) \frac{d^4 x_5}{x_{15}^2 x_{25}^2 x_{35}^2 x_{45}^2}.
\end{equation}
The property of dual conformal covariance of the integral form is not restricted to one loop but continues to all loop orders so far explored \cite{Bern:2006ew,Bern:2007ct}. For example, at three loops one of the relevant integrals requires a precise numerator factor to remain dual conformally covariant (see Fig. \ref{figure:3ladandtc}).
Note that the operation of drawing the dual graph is only possible for planar diagrams. This is the first indication that the dual conformal property is something associated with the integrability of the planar theory.

Beyond tree-level the scattering amplitudes are infrared divergent. This can be seen at the level of the integrals, e.g. as defined in (\ref{1box}). We therefore need to introduce an infrared regulator. One choice is to use dimensional regularisation. This breaks the dual conformal symmetry slightly since the integration measure is then no longer four-dimensional,
\begin{equation}
d^4x_5 \longrightarrow d^{4-2\epsilon} x_5.
\end{equation}
Alternatively one can regularise by introducing expectation values for the scalar fields \cite{Alday:2009zm}. The mass parameters then play the role of radial coordinates in $AdS_5$. This Coulomb branch approach has the advantage that the corresponding action of dual conformal symmetry transforms the regularised integral covariantly. If all the integrals appearing in the amplitude are dual conformal it implies that amplitudes on the Coulomb branch of $\mathcal{N}=4$ super Yang-Mills in the planar limit exhibit an unbroken dual conformal symmetry.
For further details on this idea see \cite{Henn:2010bk,Boels:2010mj,Henn:2010ir} and for work relating it to higher dimensional theories see \cite{Bern:2010qa,CaronHuot:2010rj,Dennen:2010dh}.

To discuss the consequences of dual conformal symmetry further it is very convenient to introduce a dual description for the scattering amplitudes. In the dual description planar amplitudes are related to Wilson loops defined on a piecewise light-like contour in the dual coordinate space. The dual conformal symmetry of the amplitude is simply the ordinary conformal symmetry of the Wilson loop. Since the conformal symmetry of a Wilson loop has a Lagrangian origin, it is possible to derive a Ward identity for it. This will show us more precisely the constraints that dual conformal symmetry places on the form of the scattering amplitudes.

\section{Amplitudes and Wilson loops}

Let us consider the general structure of a planar MHV amplitude in perturbation theory. As we have discussed we can naturally factorise MHV amplitudes into a tree-level factor and a loop-correction factor $M_n$. The factor $M_n$ contains the dependence of the amplitude on the regularisation needed to deal with the infrared divergences. Here we will use dimensional regularisation. the amplitudes will therefore depend on the regulator $\epsilon$ and some associated scale $\mu$.

Since we are discussing planar colour-ordered amplitudes it is clear that the infrared divergences will involve only a very limited dependence on the kinematical variables. Specifically, the exchange of soft or collinear gluons is limited to sectors between two adjacent incoming particles and thus the infrared divergences will factorise into pieces which depend only on a single Mandelstam variable $s_{i,i+1}=(p_i + p_{i+1})^2$.

Moreover the dependence of each of these factors is known to be of a particular exponentiated form \cite{Akhoury:1978vq,Mueller:1979ih,Collins:1980ih,Sen:1981sd,Sterman:1986aj,Botts:1989kf,Catani:1989ne,Korchemsky:1988hd,Korchemsky:1988pn,Magnea:1990zb,Korchemsky:1993uz,Catani:1998bh,Sterman:2002qn} where there is at most a double pole in the regulator in the exponent. Combining these two facts together it is most natural to write the logarithm of the loop corrections $M_n$,
\begin{equation}
{\rm log}M_n = \sum_{l=1}^{\infty} a^l \biggl[ \frac{\Gamma_{\rm cusp}^{(l)}}{(l\epsilon)^2} + \frac{\Gamma_{\rm col}^{(l)}}{l \epsilon} \biggr] \sum_i\biggl(\frac{\mu_{IR}^2}{-s_{i,i+1}}\biggr)^{l\epsilon} + F^{\rm MHV}_n(p_1,\ldots,p_n;a) + O(\epsilon).
\label{logMn}
\end{equation}
The leading infrared divergence is known to be governed by $\Gamma_{\rm cusp}(a) = \sum a^l \Gamma_{\rm cusp}^{(l)}$, the cusp anomalous dimension \cite{Ivanov:1985np,Korchemsky:1992xv}, a quantity which is so-called because it arises as the leading ultraviolet divergence of Wilson loops with light-like cusps. This is the first connection between scattering amplitudes and Wilson loops. 

In \cite{Bern:2005iz} Bern, Dixon and Smirnov (BDS) made an all order ansatz for the form of the finite part of the $n$-point MHV scattering amplitude in the planar limit. Their ansatz had the following form,
\begin{equation}
F_n^{\rm BDS}(p_1,\ldots,p_n;a) = \Gamma_{\rm cusp}(a) \mathcal{F}_n(p_1,\ldots,p_n) + c_n(a).
\label{BDSansatz}
\end{equation}
The notable feature of this ansatz is that the dependence on the coupling factorises into a single function, the cusp anomalous dimension, while the momentum dependence is contained in the coupling-independent function $\mathcal{F}_n$. The latter could therefore be defined by the one-loop amplitude, making the ansatz true by definition at one loop. The formula (\ref{BDSansatz}) was conjectured after direct calculations of the four-point amplitude to two loops \cite{Anastasiou:2003kj} and three loops \cite{Bern:2005iz}. It was found to be consistent with the five-point amplitude at two loops \cite{Cachazo:2006tj,Bern:2006vw} and three loops \cite{Spradlin:2008uu}. As we will see the results for four and five points can be explained by dual conformal symmetry. which also permits for a deviation from the form (\ref{BDSansatz}) starting from six points. Indeed the ansatz breaks down at six points and as we will see this is in agreement with dual conformal symmetry and the relation between amplitudes and Wilson loops.

In planar $\mathcal{N}=4$ super Yang-Mills theory the connection between amplitudes and Wilson loops runs deeper than just the leading infrared divergence. As we have seen one can naturally associate a collection of dual coordinates $x_i$ with a gluon scattering amplitude. Each dual coordinate is light-like separated from its neighbours,
\begin{equation}
(x_i - x_{i+1})^2 = 0
\label{lightlike}
\end{equation}
as the difference $x_i-x_{i+1}$ is the momentum $p_i$ of an on-shell massless particle. The collection of points $\{ x_i \}$ therefore naturally defines a piecewise light-like polygonal contour $C_n$ in the dual space. A natural object that one can associate with such a contour in gauge theory is the Wilson loop,
\begin{equation}
W_n = \langle \mathcal{P}exp \oint_{C_n} A \rangle.
\label{Wilsonloop}
\end{equation}
Here, in contrast to the situation for the scattering amplitude, the dual space is being treated as the actual configuration space of the gauge theory, i.e. the theory in which we compute the Wilson loop is local in this space.

A lot is known about the structure of such Wilson loops. In particular, due to the cusps on the contour at the points $x_i$ the Wilson loop is ultraviolet divergent. The divergences of such Wilson loops are intimately related to the infrared divergences of scattering amplitudes \cite{Ivanov:1985np,Korchemsky:1992xv,Korchemskaya:1996je}. Indeed the leading ultraviolet divergence is again the cusp anomalous dimension and one can write an equation very similar to that for the loop corrections to the MHV amplitude,
\begin{equation}
{\rm log}W_n = \sum_{l=1}^{\infty} a^l \biggl[ \frac{\Gamma_{\rm cusp}^{(l)}}{(l\epsilon)^2} + \frac{\Gamma^{(l)}}{l \epsilon} \biggr] \sum_i (-\mu_{UV}^2 x_{i,i+2}^2)^{l\epsilon} + F^{\rm WL}_n(x_1,\ldots,x_n;a) + O(\epsilon).
\label{logWn}
\end{equation}

The objects of most interest to us here are the two functions $F_n^{\rm MHV}$ from (\ref{logMn}) and $F_n^{\rm WL}$ from (\ref{logWn}) describing the finite parts of the amplitude and Wilson loop respectively. In fact there is by now a lot of evidence that in the planar theory, the two functions are identical up to an additive constant,
\begin{equation}
F_n^{\rm MHV}(p_1,\ldots,p_n;a) = F_n^{\rm WL}(x_1,\ldots,x_n;a) + d_n(a)
\label{MHV/WLduality}
\end{equation}
upon using the change of variables (\ref{dualcoords}).

The identification of the two finite parts was first made at strong coupling \cite{Alday:2007hr} where the AdS/CFT correspondence can be used to study the theory. In this regime the identification is a consequence of a particular T-duality transformation of the string sigma model which maps the AdS background into a dual AdS space. Shortly afterwards the identification was made in perturbation theory, suggesting that such a phenomenon is actually a non-perturbative feature. The matching was first observed at four points and one loop \cite{Drummond:2007aua} and generalised to $n$ points in \cite{Brandhuber:2007yx}. Two loop calculations then followed \cite{Drummond:2007cf,Drummond:2007au,Drummond:2007bm,Drummond:2008aq,Bern:2008ap}. In each case the duality relation (\ref{MHV/WLduality}) was indeed verified.

An important point is that dual conformal symmetry finds a natural home within the duality between amplitudes and Wilson loops. It is simply the ordinary conformal symmetry of the Wilson loop defined in the dual space. Moreover, since this symmetry is a Lagrangian symmetry from the point of view of the Wilson loop, its consequences can be derived in the form of Ward identities \cite{Drummond:2007cf,Drummond:2007au}. Importantly, conformal transformations preserve the form of the contour, i.e. light-like polygons map to light-like polygons. Thus the conformal transformations effectively act only on a finite number of points (the cusp points $x_i$) defining the contour. The generator of special conformal transformations relevant to the class of light-like polygonal Wilson loops is therefore
\begin{equation}
K_\mu = \sum_i \biggl[ x_{i\mu} x_i \cdot \frac{\partial}{\partial x_i} - \frac{1}{2} x_i^2 \frac{\partial}{\partial x_i^\mu}\biggr].
\end{equation}
Indeed the analysis of \cite{Drummond:2007au} shows that the ultraviolet divergences induce an anomalous behaviour for the finite part $F_n^{\rm WL}$ which is entirely captured by the following conformal Ward identity
\begin{equation}
K^\mu F_n^{\rm WL} = \frac{1}{2} \,\, \Gamma_{\rm cusp}(a) \sum_i (2x_i^\mu - x_{i-1}^\mu -x_{i+1}^\mu)\log x_{i-1,i+1}^2.
\label{CWI}
\end{equation}

A very important consequence of the conformal Ward identity is that the finite part of the Wilson loop is fixed up to a function of conformally invariant cross-ratios,
\begin{equation}
u_{ijkl} = \frac{x_{ij}^2 x_{kl}^2}{x_{ik}^2 x_{jl}^2}.
\end{equation}
In the cases of four and five edges, there are no such cross-ratios available due to the light-like separations of the cusp points (\ref{lightlike}). This means that the conformal Ward identity (\ref{CWI}) has a unique solution up to an additive constant.
Remarkably, the solution coincides with the Bern-Dixon-Smirnov all-order ansatz for the corresponding scattering amplitudes,
\begin{align}\label{n45}
F_4^{\rm (BDS)} &= \frac{1}{4} \Gamma_{\rm cusp}(a) \log^2 \Bigl(
\frac{x_{13}^2}{x_{24}^2} \Bigr)+ \text{ const },\\
F_5^{\rm (BDS)} &= -\frac{1}{8} \Gamma_{\rm cusp}(a) \sum_{i=1}^5 \log \Bigl(
\frac{x_{i,i+2}^2}{x_{i,i+3}^2} \Bigr) \log \Bigl(
\frac{x_{i+1,i+3}^2}{x_{i+2,i+4}^2} \Bigr) + \text{ const }. \label{n45'}
\end{align}
In fact the BDS ansatz provides a particular solution to the conformal Ward identity for any number of points. From six points onwards however the functional form is not uniquely fixed as there are conformal cross-ratios available. At six points, for example there are three of them,
\begin{equation}
u_1 = \frac{x_{13}^2 x_{46}^2}{x_{14}^2 x_{36}^2}, \qquad u_2 = \frac{x_{24}^2 x_{51}^2}{x_{25}^2 x_{41}^2}, \qquad u_3 = \frac{x_{35}^2 x_{62}^2}{x_{36}^2 x_{41}^2}.
\end{equation}
The solution to the Ward identity is therefore
\begin{equation}\label{solutionWI}
F_6^{\rm (WL)} = F_6^{\rm (BDS)} + f(u_1,u_2,u_3;a)\ .
\end{equation}
Here, upon the identification $p_i=x_i-x_{i+1}$,
\begin{align}\label{BDS6point}
  F_{6} ^{\rm (BDS)}   =  \frac{1}{4} \Gamma_{\rm
 cusp}(a)\sum_{i=1}^{6} &\bigg[
    - \log \Bigl(
\frac{x_{i,i+2}^2}{x_{i,i+3}^2} \Bigr)\log\Bigl(
\frac{x_{i+1,i+3}^2}{x_{i,i+3}^2} \Bigr)
\nonumber\\
 &  +\frac{1}{4} \log^2 \Bigl( \frac{x_{i,i+3}^2}{x_{i+1,i+4}^2}
\Bigr)   -\frac{1}{2} {\rm{Li}}_{2}\Bigl(1-
 \frac{x_{i,i+2}^2  x_{i+3,i+5}^2}{x_{i,i+3}^2 x_{i+2,i+5}^2}
\Bigr) \bigg] + \text{ const }\,,
\end{align}
while $f(u_1,u_2,u_3;a)$ is some function of the three
cross-ratios and the coupling. As we have discussed the function $f$ is not fixed by the Ward identity and has to be determined by explicit calculation of the Wilson loop. The calculation of \cite{Brandhuber:2007yx} shows that at one loop $f$ vanishes (recall that at one loop the BDS ansatz is true by definition and the Wilson loop and MHV amplitude are known to agree for an arbitrary number of points). At two loops, direct calculation shows that $f$ is non-zero \cite{Drummond:2007bm,Drummond:2008aq}. Moreover the calculation \cite{Bern:2008ap} of the six-particle MHV amplitude shows explicitly that the BDS ansatz breaks down at two loops and the same function appears there,
\begin{equation}
F_6^{\rm MHV} = F_6^{\rm WL} + const,\qquad F_6^{\rm MHV} \neq F_6^{\rm BDS}.
\end{equation}
The agreement between the two functions $F_6^{\rm MHV}$ and $F_6^{\rm WL}$ was verified numerically to within the available accuracy. Subsequently the integrals appearing in the calculation of the finite part of the of the hexagonal Wilson loop have been evaluated analytically in terms of multiple polylogarithms \cite{DelDuca:2010zg}.

Further calculations of polygonal Wilson loops have been performed. The two-loop diagrams appearing for an arbitrary number of points have been described in \cite{Anastasiou:2009kna} where numerical evaluations of the seven-sided and eight-sided light-like Wilson loops were made. These functions have not yet been compared with the corresponding MHV amplitude calculations \cite{Vergu:2009zm,Vergu:2009tu} due the difficulty of numerically evaluating the relevant integrals. However given the above evidence it seems very likely that the agreement between MHV amplitudes and light-like polygonal Wilson loops will continue to an arbitrary number of points, to all orders in the coupling.

While the agreement between Wilson loops is fascinating it is clearly not the end of the story. Firstly the duality as we have described it applies only to the MHV amplitudes. In the strict strong coupling limit this does not matter since all amplitudes are dominated by the minimal surface in AdS, independently of the helicity configuration \cite{Alday:2007hr}. At weak coupling that is certainly not the case and non-MHV amplitudes reveal a much richer structure than their MHV counterparts. Recently the duality has been extended to take into account non-MHV amplitudes \cite{Mason:2010yk,CaronHuot:2010ek} by introducing an appropriate supersymmetrisation of the Wilson loop.

Even without regard to a dual Wilson loop, one may still ask what happens to dual conformal symmetry for non-MHV amplitudes. To properly ask this question one must first deal with the notion of helicity since non-MHV amplitudes are not naturally written as a product of tree-level and loop-correction contributions. In considering different helicity configurations we are led to the notion of dual superconformal symmetry.

\section{Superconformal and dual superconformal symmetry}

The on-shell supermultiplet of $\mathcal{N}=4$ super Yang-Mills theory is conveniently described by a superfield $\Phi$, dependent on Grassmann parameters $\eta^A$ which transform in the fundamental representation of $su(4)$. The on-shell superfield can be expanded as follows
\begin{equation}
\Phi = G^+ + \eta^A \Gamma_A + \tfrac{1}{2!} \eta^A \eta^B S_{AB} + \tfrac{1}{3!} \eta^A \eta^B \eta^C \epsilon_{ABCD} \overline{\Gamma}^D + \tfrac{1}{4!} \eta^A \eta^B \eta^C \eta^D \epsilon_{ABCD} G^-.
\label{onshellmultiplet}
\end{equation}
Here $G^+,\Gamma_A,S_{AB}=\tfrac{1}{2}\epsilon_{ABCD}\overline{S}^{CD},\overline{\Gamma}^A,G^-$ are the positive helicity gluon, gluino, scalar, anti-gluino and negative helicity gluon states respectively. Each of the possible states $\phi \in \{G^+,\Gamma_A,S_{AB},\overline{\Gamma}^A,G^-\}$ carries a definite on-shell momentum
\begin{equation}
p^{\alpha \dot\alpha} = \lambda^\alpha \tilde{\lambda}^{\dot \alpha},
\end{equation}
and a definite weight $h$ (called helicity) under the rescaling
\begin{equation}
\lambda \longrightarrow \alpha \lambda, \qquad \tilde{\lambda} \longrightarrow \alpha^{-1} \tilde{\lambda}, \qquad \phi(\lambda,\tilde{\lambda}) \longrightarrow \alpha^{-2h} \phi(\lambda,\tilde{\lambda}).
\end{equation}
The helicities of the states appearing in (\ref{onshellmultiplet}) are $\{+1,+\tfrac{1}{2},0,-\tfrac{1}{2},-1\}$ respectively. If, in addition, we assign $\eta$ to transform in the same way as $\tilde{\lambda}$,
\begin{equation}
\eta^A \longrightarrow \alpha^{-1} \eta^A,
\end{equation}
then the whole superfield $\Phi$ has helicity 1. In other words the helicity generator\footnote{In terms of the superconformal algebra $su(2,2|4)$, the operator $h$ is the central charge.},
\begin{equation}
h = -\tfrac{1}{2} \lambda^\alpha \frac{\partial}{\partial \lambda^\alpha} + \tfrac{1}{2} \tilde{\lambda}^{\dot \alpha} \frac{\partial}{\partial \tilde{\lambda}^{\dot \alpha}} + \tfrac{1}{2} \eta^A \frac{\partial}{\partial \eta^A},
\end{equation}
acts on $\Phi$ in the following way,
\begin{equation}
h \Phi = \Phi.
\end{equation}
When we consider superamplitudes, i.e. colour-ordered scattering amplitudes of the on-shell superfields, then the helicity condition (or `homogeneity condition') is satisfied for each particle, 
\begin{equation}
h_i \mathcal{A}(\Phi_1,\ldots,\Phi_n) = \mathcal{A}(\Phi_1,\ldots,\Phi_n), \qquad i=1,\ldots,n.
\label{Ahelicity}
\end{equation}
The tree-level amplitudes in $\mathcal{N}=4$ super Yang-Mills theory can be written as follows,
\begin{equation}
\mathcal{A}(\Phi_1,\ldots,\Phi_n)=\mathcal{A}_n =  \frac{\delta^4(p) \delta^8(q)}{\langle12\rangle \ldots \langle n1\rangle } \mathcal{P}_n(\lambda_i,\tilde{\lambda}_i,\eta_i) = \mathcal{A}_n^{\rm MHV} \mathcal{P}_n.
\label{amp}
\end{equation}
The MHV tree-level amplitude,
\begin{equation}
\mathcal{A}_n^{\rm MHV} = \frac{\delta^4(p) \delta^8(q)}{\langle 12 \rangle \ldots \langle n1 \rangle},
\end{equation}
contains the delta functions $\delta^4(p) \delta^8(q)$ which are a consequence of translation invariance and supersymmetry and it can be factored out leaving behind a function with no helicity,
\begin{equation}
h_i \mathcal{P}_n = 0, \qquad i=1,\ldots,n.
\label{Phelicity}
\end{equation}
The function $\mathcal{P}_n$ can be expanded in terms of increasing Grassmann degree (the Grassmann degree always comes in multiples of 4 dues to invariance under $su(4)$),
\begin{equation}
\mathcal{P}_n = 1 + \mathcal{P}_n^{\rm NMHV} + \mathcal{P}_n^{\rm NNMHV} + \,\, \ldots \,\,+ \mathcal{P}_n^{\overline{\rm MHV}}.
\end{equation}
The explicit form of the function $\mathcal{P}_n$ which encodes all tree-level amplitudes was found in \cite{Drummond:2008cr} by solving a supersymmetrised version \cite{Brandhuber:2008pf,ArkaniHamed:2008gz,Elvang:2008na} of the BCFW recursion relations \cite{Britto:2004ap,Britto:2005fq}.

Maximally supersymmetric Yang-Mills is a superconformal field theory so we should expect that this is reflected in the structure of the scattering amplitudes. Indeed the space of functions of the variables $\{\lambda_i,\tilde{\lambda}_i,\eta_i\}$ admits a representation of the superconformal algebra. 
The explicit form of the representation acting on the on-shell superspace coordinates $(\lambda_i,\tilde{\lambda}_i,\eta_i)$ is essentially the oscillator representation \cite{Witten:2003nn}. 
\begin{align}
& p^{\dot{\alpha}\alpha }  =  \sum_i \tilde{\lambda}_i^{\dot{\alpha}}\lambda_i^{\alpha} \,, & &
k_{\alpha \dot{\alpha}} = \sum_i \partial_{i \alpha} \partial_{i \dot{\alpha}} \,,\notag\\
&\overline{m}_{\dot{\alpha} \dot{\beta}} = \sum_i \tilde{\lambda}_{i (\dot{\alpha}} \partial_{i
\dot{\beta} )}, & & m_{\alpha \beta} = \sum_i \lambda_{i (\alpha} \partial_{i \beta )}
\,,\notag\\
& d =  \sum_i [\tfrac{1}{2}\lambda_i^{\alpha} \partial_{i \alpha} +\tfrac{1}{2}
\tilde{\lambda}_i^{\dot{\alpha}} \partial_{i
    \dot{\alpha}} +1], & & r^{A}{}_{B} = \sum_i [-\eta_i^A \partial_{i B} + \tfrac{1}{4}\delta^A_B \eta_i^C \partial_{i C}]\,,\notag\\
&q^{\alpha A} =  \sum_i \lambda_i^{\alpha} \eta_i^A \,, &&   \bar{q}^{\dot\alpha}_A
= \sum_i \tilde\lambda_i^{\dot \alpha} \partial_{i A} \,, \notag\\
& s_{\alpha A} =  \sum_i \partial_{i \alpha} \partial_{i A}, & &
\bar{s}_{\dot\alpha}^A = \sum_i \eta_i^A \partial_{i \dot\alpha}\,,\notag\\
&c = \sum_i [1 + \tfrac{1}{2} \lambda_i^{\alpha} \partial_{i \alpha} - \tfrac{1}{2} \tilde\lambda^{\dot \alpha}_i \partial_{i \dot \alpha} - \tfrac{1}{2} \eta^A_i \partial_{iA} ]\,.
\end{align}
This realisation of the superconformal algebra also appears in the discussion of gauge-invariant operators \cite{chapSuperconf}.
From the commutation relations of the superconformal algebra one finds that the algebra is generically $su(2,2|4)$ with central charge $c = \sum_i (1-h_i)$. Amplitudes are in the space of functions with helicity 1 for each particle so we have that $c=0$ after imposing the helicity conditions (\ref{Ahelicity}) and the algebra acting on the space of homogeneous functions is $psu(2,2|4)$.

At tree-level there are no infrared divergences and amplitudes are annihilated by the generators of the standard superconformal symmetry (up to contact terms which vanish for generic configurations of the external momenta, see \cite{Bargheer:2009qu,Korchemsky:2009hm,Sever:2009aa}),
\begin{equation}
j_a \mathcal{A}_n = 0. \label{scs}
\end{equation}
Here we use the notation $j_a$ for any generator of the superconformal algebra $psu(2,2|4)$,
\begin{equation}
j_a \in \{p^{\alpha \dot \alpha},q^{\alpha A}, \bar{q}^{\dot \alpha}_A,m_{\alpha \beta}, \bar{m}_{\dot \alpha \dot \beta},r^A{}_B,d,s^\alpha_A,\bar{s}_{\dot \alpha}^A,k_{\alpha \dot \alpha} \}.
\end{equation}

As well as superconformal symmetry one can naturally define the action of dual superconformal symmetry \cite{Drummond:2008vq} on colour-ordered amplitudes. We have already seen that one can define dual coordinates $x_i$ related to the particle momenta. These variables implicitly solve the momentum conservation condition imposed by the delta function $\delta^4(p)$. The presence of a corresponding $\delta^8(q)$ due to supersymmetry suggests defining new fermionic dual variables $\theta_i$ related to the supercharges. Together these variables parametrise a chiral dual superspace,
\begin{equation}
x_i^{\alpha \dot \alpha} - x_{i+1}^{\alpha \dot\alpha} = \lambda_i^\alpha \tilde\lambda_i^{\dot \alpha}=p_i^{\alpha \dot\alpha}, \qquad \theta_i^{\alpha A} - \theta_{i+1}^{\alpha A} = \lambda_i^\alpha \eta_i^A = q_i^{\alpha a}.
\label{dualvars}
\end{equation}
Dual superconformal symmetry acts canonically on the dual superspace variables $x_i,\theta_i$. It also acts on the on-shell superspace variables $\{ \lambda_i ,\tilde{\lambda}_i, \eta_i \}$ in order to be compatible with the defining relations (\ref{dualvars}). 
In particular one can deduce an action of dual conformal inversions on $\lambda_i,\tilde{\lambda}_i$ compatible with (\ref{dualvars}),
\begin{equation}
I[\lambda_i^{\alpha}] = \frac{(\lambda_i x_i)^{\dot\alpha}}{x_i^2}, \qquad I[\tilde{\lambda}_i^{\dot\alpha}] = \frac{(\tilde{\lambda}_i x_{i})^{\alpha}}{x_{i+1}^2}.
\label{dualinversion}
\end{equation}
Alternatively one may think about infinitesimal dual superconformal transformations. In this case one should extend the canonical generators on the chiral superspace variables $x_i$ and $\theta_i$ to act on the on-shell superspace variables $\lambda_i,\tilde{\lambda}_i$ and $\eta_i$ so that they commute with the constraints (\ref{dualvars}) modulo the constraints themselves. We give explicitly the form of the dual conformal generator,
\begin{align}
K_{\alpha \dot{\alpha}} = \sum_i [&x_{i \alpha}{}^{\dot{\beta}} x_{i
    \dot{\alpha}}{}^{\beta} \partial_{i \beta \dot{\beta}} + x_{i
    \dot{\alpha}}{}^{\beta} \theta_{i \alpha}^B \partial_{i \beta B} \notag \\
    &+
  x_{i \dot{\alpha}}{}^{\beta} \lambda_{i \alpha} \partial_{i \beta}
  + x_{i+1 \,\alpha}{}^{\dot{\beta}} \tilde{\lambda}_{i \dot{\alpha}}
  \partial_{i \dot{\beta}} + \tilde{\lambda}_{i \dot{\alpha}} \theta_{i+1\,
    \alpha}^B \partial_{i B}]\,.
\label{dualsc}
\end{align}
The anti-chiral fermionic generators are also of interest,
\begin{equation}
\overline{Q}_{\dot{\alpha}}^A = \sum_i [\theta_i^{\alpha A}
  \partial_{i \alpha \dot{\alpha}} + \eta_i^A \partial_{i \dot{\alpha}}], \qquad \overline{S}_{\dot{\alpha} A} = \sum_i [x_{i \dot{\alpha}}{}^{\beta}
  \partial_{i \beta A} + \tilde{\lambda}_{i \dot{\alpha}}
  \partial_{iA}]\,.
\end{equation}
The remaining generators can be found in \cite{Drummond:2008vq}. Note that when restricted to the on-shell superspace, the generators $\bar{Q}_{\dot\alpha}^A$ and $\bar{S}_{\dot \alpha A}$ coincide with the generators $\bar{s}_{\dot\alpha}^A$ and $\bar{q}_{\dot \alpha A}$ respectively from the original superconformal algebra.

Now that the symmetry has been defined we must also specify how the amplitudes transform. In \cite{Drummond:2008vq}
 it was conjectured, based on the supersymmetric forms of the MHV and next-to-MHV (NMHV) tree-level amplitudes, that the full tree-level superamplitude $\mathcal{A}_{n,{\rm tree}}$ is covariant under dual superconformal symmetry. Explicitly, it was conjectured that the tree-level amplitudes obey
\begin{align}
K^{\alpha \dot \alpha} \mathcal{A}_n &= -\sum_i x_i^{\alpha \dot \alpha} \mathcal{A}_n,
\label{dualKcov}\\
S^{\alpha A} \mathcal{A}_n &= -\sum_i \theta_i^{\alpha A} \mathcal{A}_n,
\label{dualScov}
\end{align}
together with the obvious properties $D \mathcal{A}_n = n \mathcal{A}_n$ and $C \mathcal{A}_n = n \mathcal{A}_n$. The remaining generators of the dual superconformal algebra annihilate the amplitudes.

The amplitudes can be expressed in the dual variables by eliminating ($\tilde{\lambda}_i,\eta_i$) in favour of $(x_i,\theta_i)$. If we relax the cyclicity condition on the dual points so that $x_1 \neq x_{n+1}$ and $\theta_1 \neq \theta_{n+1}$ then we have
\begin{equation}
\mathcal{A}_n = \frac{\delta^4(x_1-x_{n+1}) \delta^8(\theta_1-\theta_{n+1})}{\langle 12\rangle \ldots \langle n1\rangle } \mathcal{P}_n(x_i,\theta_i).
\end{equation}
From the dual conformal transformations described earlier we can see that the MHV prefactor itself satisfies the covariance conditions (\ref{dualKcov},\ref{dualScov}). The function $\mathcal{P}_n$ must therefore be dual superconformally invariant.
At the MHV level the function $\mathcal{P}_n$ is simply 1 while at NMHV level it is given by \cite{Drummond:2008vq,Drummond:2008bq,Drummond:2008cr}
\begin{equation}
\mathcal{P}_n^{\rm NMHV} = \sum_{a,b} R_{n,ab} 
\end{equation}
where the sum runs over the range $2 \leq a < b \leq n-1$ (with $a$ and $b$ separated by at least two) and
\begin{equation} 
R_{n,ab} = \frac{\langle a\,a-1 \rangle \langle b \, b-1 \rangle\delta^4\bigl(\langle n | x_{na}x_{ab}|\theta_{bn}\rangle + \langle n | x_{nb}x_{ba}|\theta_{an}\rangle\bigr)}{x_{ab}^2 \langle n |x_{na}x_{ab}|b\rangle \langle n |x_{na}x_{ab}|b-1\rangle \langle n |x_{nb}x_{ba}|a\rangle \langle n |x_{nb}x_{ba}|a-1\rangle}.
\end{equation}
This formula was originally constructed in \cite{Drummond:2008vq} by supersymmetrising the three-mass coefficients of NMHV gluon scattering amplitudes at one loop in \cite{Bern:2004bt}. It was then derived from supersymmetric generalised unitarity \cite{Drummond:2008bq} as the general form of the one-loop three-mass box coefficient. One can see from the transformations described earlier that each $R_{n,ab}$ is by itself a dual superconformal invariant.

The pattern of invariance continues for all tree-level amplitudes. Indeed the conjecture (\ref{dualKcov},\ref{dualScov}) was shown to hold recursively in \cite{Brandhuber:2008pf}, using the supersymmetric BCFW recursion relations. Indeed the BCFW recursion relations admit a closed form solution for all tree-level amplitudes in $\mathcal{N}=4$ super Yang-Mills theory \cite{Drummond:2008cr} with each term being a dual superconformal invariant by itself. 
 
What happens to the symmetry of scattering amplitudes beyond tree-level? Firstly we expect a breakdown of the original conformal symmetry due to infrared divergences. One might also expect a breakdown of the dual superconformal symmetry in the same way. However, at least for the MHV amplitudes we have already seen that the dual conformal symmetry is broken only mildly in that it is controlled by the anomalous Ward identity (\ref{CWI}). Based on analysis of the one-loop NMHV amplitudes it was conjectured in \cite{Drummond:2008vq} that the same anomaly controls the breakdown for all amplitudes, irrespective of the MHV degree. Specifically if one writes the all-order superamplitude as a product of the MHV superamplitude and an infrared finite ratio function\footnote{The ratio function $R_n$ is infrared finite because the infrared divergences of all planar amplitudes are independent of the helicity configuration and are thus contained entirely in the factor $\mathcal{A}_n^{\rm MHV}$.},
\begin{equation}
\mathcal{A}_n = \mathcal{A}_n^{\rm MHV} R_n,
\end{equation}
then the conjecture states that, setting the regulator to zero, $R_n$ is dual conformally invariant,
\begin{equation}
K^\mu R_n = 0.
\label{nonMHVinvariance}
\end{equation}

In \cite{Drummond:2008bq} it was argued that this conjecture holds for NMHV amplitudes at one loop, based on explicit calculations up to nine points using supersymmetric generalised unitarity. Subsequently \cite{Brandhuber:2009kh} it has been argued to hold for all one-loop amplitudes by analysing the dual conformal anomaly arising from infrared divergent two-particle cuts.
 
Note that the conjecture (\ref{nonMHVinvariance}) makes reference only to the dual conformal generator $K$ and not to the full set of dual superconformal transformations. The reason is that some of these transformations overlap with the broken part of the original superconformal symmetry. In particular the generator $\bar{Q}$ is not a symmetry of the ratio function $R_n$. This fact is related to the breaking of the original superconformal invariance by loop corrections since $\bar{Q}$ is really the same symmetry as $\bar{s}$. Indeed, even at tree-level $\bar{s}$ is subtly broken by contact term contributions \cite{Bargheer:2009qu,Korchemsky:2009hm,Sever:2009aa}. At one loop unitarity relates the discontinuity of the amplitude in a particular channel to the product of two tree-level amplitudes integrated over the allowed phase space of the exchanged particles. The subtle non-invariance of the trees therefore translates into non-invariance of the discontinuity and therefore of the loop amplitude itself \cite{Korchemsky:2009hm,Sever:2009aa,Beisert:2010gn}. In \cite{Beisert:2010gn} a deformation of the ordinary and dual superconformal generators is presented which takes into account the one-loop corrections to the amplitudes. The existence of Wilson loops which take into account the non-MHV amplitudes \cite{Mason:2010yk,CaronHuot:2010ek} suggests that the universality of the dual conformal anomaly is very natural from the dual perspective.

\section{Yangian symmetry}

In order to put the dual superconformal symmetry on the same footing as invariance under the standard superconformal algebra (\ref{scs}), the covariance (\ref{dualKcov},\ref{dualScov}) can be rephrased as an invariance of $\mathcal{A}_n$ by a simple redefinition of the generators \cite{Drummond:2009fd},
\begin{align}
K'^{\alpha \dot{\alpha}} &= K^{\alpha \dot{\alpha}} + \sum_i x_i^{\alpha \dot{\alpha}}, \\
S'^{\alpha A} &= S^{\alpha A} + \sum_i \theta_i^{\alpha A}, \\
D' &= D - n.
\end{align}
The redefined generators still satisfy the commutation relations of the superconformal algebra, but now with vanishing central charge, $C' = 0$.
Then dual superconformal symmetry is simply
\begin{equation}
J'_a \mathcal{A}_n = 0.
\end{equation}
Here we use the notation $J'_a$ for any generator of the {\sl dual} copy of $psu(2,2|4)$,
\begin{equation}
J'_a \in \{P_{\alpha \dot\alpha},Q_{\alpha A}, \bar{Q}_{\dot\alpha}^A,M_{\alpha \beta}, \overline{M}_{\dot{\alpha} \dot{\beta}},R^A{}_B,D',S_\alpha^{\prime A},\overline{S}^{\dot{\alpha}}_A,K'^{\alpha \dot\alpha} \}.
\end{equation}

In order to have both symmetries acting on the same space it is useful to restrict the dual superconformal generators to act only on the on-shell superspace variables $(\lambda_i,\tilde{\lambda}_i,\eta_i)$. Then one finds that the generators $P_{\alpha \dot\alpha},Q_{\alpha A}$ become trivial while the generators $\{\bar{Q},M,\bar{M},R,D',\bar{S}\}$ coincide (up to signs) with generators of the standard superconformal symmetry. The non-trivial generators which are not part of the $j_a$ are $K'$ and $S'$. In \cite{Drummond:2009fd} it was shown that the generators $j_a$ and $S'$ (or $K'$) together generate the Yangian of the superconformal algebra, $Y(psu(2,2|4))$. The generators $j_a$ form the level-zero $psu(2,2|4)$ subalgebra\footnote{We use the symbol $[O_1,O_2]$ to denote the bracket of the Lie superalgebra, $[O_2,O_1] = (-1)^{1+|O_1||O_2|}[O_1,O_2]$.},
\begin{equation}
[j_a,j_b] = f_{ab}{}^{c} j_c.
\end{equation}
In addition there are level-one generators $j_a^{(1)}$ which transform in the adjoint under the level-zero generators,
\begin{equation}
[j_a,j_b\!{}^{(1)}] = f_{ab}{}^{c} j_c\!{}^{(1)}.
\end{equation}
Higher commutators among the generators are constrained by the Serre relation\footnote{The symbol $\{\cdot,\cdot,\cdot\}$ denotes the graded symmetriser.},
\begin{align}
&[j^{(1)}_a , [j^{(1)}_b,j_c]] + (-1)^{|a|(|b| + |c|)} [j^{(1)}_b,[j^{(1)}_c,j_a]] + (-1)^{|c|(|a|+|b|)} [j^{(1)}_c,[j^{(1)}_a,j_b]] \notag \\
&= h^2 (-1)^{|r||m|+|t||n|}\{j_l,j_m,j_n\} f_{ar}{}^{l} f_{bs}{}^{m} f_{ct}{}^{n} f^{rst}.
\end{align}
The level-zero generators are represented by a sum over single particle generators,
\begin{equation}
j_a = \sum_{k=1}^n j_{ka}.
\label{levelzero}
\end{equation}
The level-one generators are represented by the bilocal formula \cite{Dolan:2003uh,Dolan:2004ps},
\begin{equation}
j_a\!{}^{(1)} = f_{a}{}^{cb} \sum_{k<k'} j_{kb} j_{k'c}.
\label{bilocal}
\end{equation}
Thus finally the full symmetry of the tree-level amplitudes can be rephrased as
\begin{equation}
y \mathcal{A}_n = 0,
\end{equation}
for any $y \in Y(psu(2,2|4))$.

It is particularly simple to describe the symmetry in terms of twistor variables. These variables will become especially relevant in the next section where we relate the symmetry to a conjectured formula for all leading singularities of planar $\mathcal{N}=4$ SYM amplitudes.
In $(2,2)$ signature the twistor variables are simply related to the on-shell superspace variables $(\lambda,\tilde{\lambda},\eta)$ by a Fourier transformation $\lambda \longrightarrow \tilde{\mu}$.
Expressed in terms of the twistor space variables $\mathcal{Z}^{\AA} = (\tilde{\mu}^{\alpha}, \tilde\lambda^{\dot\alpha} , \eta^A)$, the level-zero and level-one generators of the Yangian symmetry assume a simple form
\begin{align}
j^{\AA}{}_{\BB} &= \sum_i \cZ_i^{\AA} \frac{\partial}{\partial \cZ_i^{\BB}}, \label{twistorsconf}\\
j^{(1)}{}^{\AA}{}_{\BB} &= \sum_{i<j} (-1)^{\CC}\Bigl[\cZ_i^{\AA} \frac{\partial}{\partial \cZ_i^{\CC}} \cZ_j^{\CC} \frac{\partial}{\partial \cZ_j^{\BB}} - (i,j) \Bigr].
\label{twistoryangian}
\end{align}
Both of the formulas (\ref{twistorsconf}) and (\ref{twistoryangian}) are understood to have the supertrace proportional to $(-1)^{\AA} \delta^{\AA}_{\BB}$ removed\footnote{One removes the supertrace of an $(m|m)\times(m|m)$ matrix $M^{\AA}{}_{\BB}$ by forming the combination $M^{\AA}{}_{\BB}-\tfrac{1}{2m} (-1)^{\AA+\CC} \delta^{\AA}_{\BB}M^{\CC}{}_{\CC}$. In addition to the supertrace $gl(m|m)$ also has a central element proportional to the identity $\delta^{\AA}_{\BB}$. In the present context the trace of (\ref{twistorsconf}) vanishes due to the homogeneity conditions while (\ref{twistoryangian}) is traceless due to the antisymmetrisation in $i$ and $j$.}. In this representation the generators of superconformal symmetry are first-order operators while the level-one Yangian generators are second order.

In \cite{Drummond:2010qh} it was demonstrated that there exists an alternative T-dual representation of the symmetry. The dual superconformal symmetries $J_a$ which play the role of the level-zero generators, while some of the level-one generators are induced by the ordinary superconformal symmetries.
In this case, the generators act on the function $\mathcal{P}_n$, where the MHV tree-level amplitude is factored out. 
\begin{equation}
J_a \mathcal{P}_n = 0, \qquad J^{(1)}_a \mathcal{P}_n = 0.
\end{equation}
It is possible to rewrite the generators in the momentum (super)twistor representation defined in \cite{Hodges:2009hk} $\cW_i^{\AA} = (\lambda_i^\alpha,\mu_i^{\dot \alpha},\chi_i^A)$. These variables are algebraically related to the on-shell superspace variables $(\lambda,\tilde\lambda,\eta)$ via the introduction of dual coordinates (\ref{dualvars})
and are the twistors associated to this dual coordinate space,
\begin{equation}
\mu_i^{\dot\alpha} = x_i^{\alpha \dot\alpha} \lambda_{i \alpha}, \qquad \chi_i^A = \theta_i^{\alpha A} \lambda_{i \alpha}.
\end{equation}
These variables linearise dual superconformal symmetry in complete analogy with the twistor variables $\mathcal{Z}_i$ and the original superconformal symmetry,
\begin{equation}
J^{\AA}{}_{\BB} = \sum_i \mathcal{W}_i^{\AA} \frac{\partial}{\partial \mathcal{W}_i^{\BB}}.
\label{dualsconfmomtwistor}
\end{equation}
The original conformal invariance of the amplitude $k_{\alpha \dot\alpha} \mathcal{A}_n=0$ induces a second-order operator which annihilates $\mathcal{P}_n$. When combined with the dual superconformal symmetry one finds that the following second-order operators annihilate $\mathcal{P}_n$,
\begin{equation}
J^{(1)}{}^{\AA}{}_{\BB} = \sum_{i<j} (-1)^{\CC} \biggl[ \mathcal{W}_i^{\AA} \frac{\partial}{\partial \mathcal{W}_i^{\CC}} \mathcal{W}_j^{\CC} \frac{\partial}{\partial \mathcal{W}_j^{\BB}} - (i,j)\biggr].
\label{levelonemomtwistor}
\end{equation}
As in the case of the original superconformal symmetry, both formulas (\ref{dualsconfmomtwistor}) and (\ref{levelonemomtwistor}) are understood to have the supertrace removed.

\begin{figure}
\ifarxiv
\centerline{\includegraphics[height=8cm]{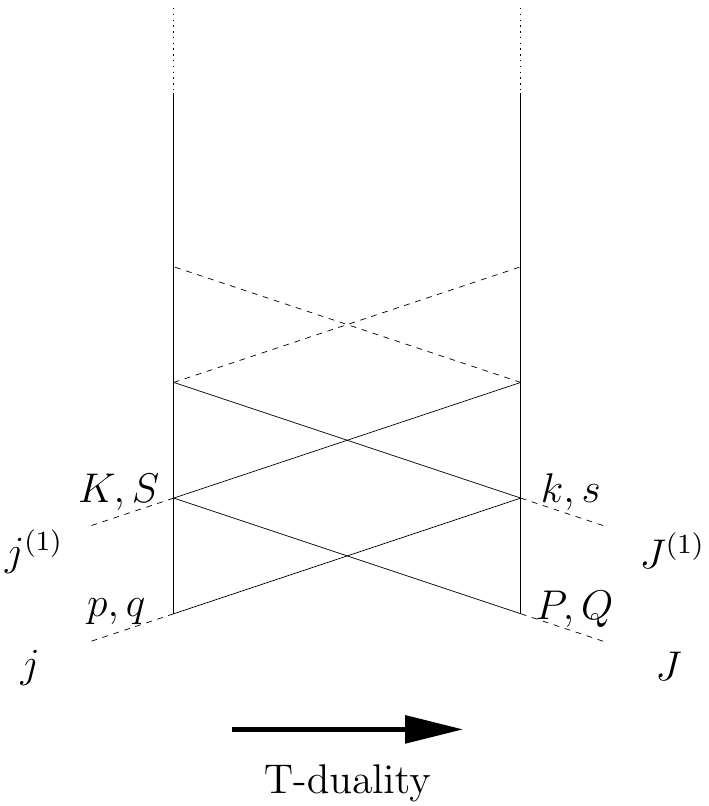}} 
\else
\centerline{\includegraphics[height=8cm]{tduality-revtimes}} 
\fi
\caption{The tower of symmetries acting on scattering amplitudes in $\mathcal{N}=4$ super Yang-Mills theory. The original superconformal charges are denoted by $j$ and the dual ones by $J$. Each can be thought of as the level-zero part of the Yangian $Y(psu(2,2|4))$. The dual superconformal charges $K$ and $S$ form part of the level-one $j^{(1)}$ while the original superconformal charges $k$ and $s$ form part of the level one charges $J^{(1)}$. In each representation the `negative' level ($P$ and $Q$ or $p$ and $q$) is trivialised. T-duality maps $j$ to $J$ and $j^{(1)}$ to $J^{(1)}$.}
\label{Fig:T-duality}
\end{figure}

The operation we have performed is summarised in Fig. \ref{Fig:T-duality}. A very similar picture also arises in considering the combined action of bosonic and fermionic T-duality in the AdS sigma model \cite{Berkovits:2008ic,Beisert:2008iq,Beisert:2009cs}.
It can be thought of as the algebraic expression of T-duality in the perturbative regime of the theory. 

Having described the symmetry of the theory, one might naturally ask how one can produce invariants. This question has been addressed in various papers \cite{Korchemsky:2009jv,Drummond:2010qh,Drummond:2010uq,Korchemsky:2010ut}. It turns out to be intimately connected to another conjecture about the leading singularities of the scattering amplitudes of $\mathcal{N}=4$ super Yang-Mills theory.

\section{Grassmannian formulas}

In \cite{ArkaniHamed:2009dn} a remarkable formula was proposed which computes leading singularities of scattering amplitudes in $\mathcal{N}=4$ super Yang-Mills theory. The formula takes the form of an integral over the Grassmannian $G(k,n)$, the space of complex $k$-planes in $\mathbb{C}^n$. The integrand is a specific $k(n-k)$-form $K$ to be integrated over cycles $C$ of the corresponding dimension, with the integral being treated as a multi-dimensional contour integral.
The result obtained depends on the choice of contour and is non-vanishing for closed contours because the form has poles located on certain hyperplanes in the Grassmannian,
\begin{equation}
\mathcal{L} = \int_C K.
\end{equation}

The form $K$ is constructed from a product of superconformally-invariant delta functions
of linear combinations of twistor variables. It is through this factor that the integral depends on the kinematic data of the $n$-point scattering amplitude of the gauge theory. The delta functions are multiplied by a cyclically invariant function on the Grassmannian which has poles. Specifically the formula takes the following form in twistor space
\begin{equation}
\mathcal{L}_{\rm ACCK}(\cZ) = \int \frac{D^{k(n-k)}c }{\mathcal{M}_1 \ldots \mathcal{M}_n} \prod_{a=1}^k \delta^{4|4}\Bigl(\sum_{i=1}^n c_{ai} \cZ_i\Bigr) ,
\label{ACCK}
\end{equation}
where the $c_{ai}$ are complex parameters which are integrated choosing a specific contour. The form $D^{k(n-k)}c$ is the natural holomorphic globally $gl(n)$-invariant and locally $sl(k)$-invariant $(k(n-k),0)$-form given explicitly in \cite{Mason:2009qx}.
The denominator is the cyclic product of consecutive $(k \times k)$ minors $\mathcal{M}_p$  made from the columns $
p,\ldots,p+k-1$ of the $(k \times n)$ matrix of the $c_{ai}$
\begin{equation}
\mathcal{M}_{p} \equiv (p~p+1~p+2\ldots p+k-1)  .
\end{equation}
As described in \cite{ArkaniHamed:2009dn} the formula (\ref{ACCK}) has a $GL(k)$ gauge symmetry which implies that $k^2$ of the $c_{ai}$ are gauge degrees of freedom and therefore should not be integrated over. The remaining $k(n-k)$ are the true coordinates on the Grassmannian. 
This formula (\ref{ACCK}) produces leading singularities of N${}^{k-2}$MHV scattering amplitudes when suitable closed integration contours are chosen. This fact was explicitly verified up to eight points in \cite{ArkaniHamed:2009dn} and it was conjectured that the formula produces all possible leading singularities at all orders in the perturbative expansion.

The formula (\ref{ACCK}) has a T-dual version \cite{Mason:2009qx}, expressed in terms of momentum twistors. The momentum twistor Grassmannian formula takes the same form as the original
\begin{equation}
\mathcal{L}_{\rm MS}(\mathcal{W}) = \int \frac{D^{k(n-k)}t}{\mathcal{M}_1 \ldots \mathcal{M}_n} \prod_{a=1}^k  \delta^{4|4}\Bigl(\sum_{i=1}^n t_{ai} \cW_i\Bigr) ,
\label{MS}
\end{equation}
but now it is the dual superconformal symmetry that is manifest. The integration variables $t_{ai}$ are again a $(k \times n)$ matrix of complex parameters and we use the notation $\mathcal{M}_p$ to refer to $(k \times k)$ minors made from the matrix of the $t_{ai}$. The formula (\ref{MS}) produces the same objects as (\ref{ACCK}) but now with the MHV tree-level amplitude factored out. They therefore contribute to N${}^k$MHV amplitudes.

The equivalence of the two formulations (\ref{ACCK}) and (\ref{MS}) was shown in \cite{ArkaniHamed:2009vw} via a change of variables. Therefore, since each of the formulas has a different superconformal symmetry manifest, they both possess an invariance under the Yangian $Y(psl(4|4))$. The Yangian symmetry of these formulas was explicitly demonstrated in \cite{Drummond:2010qh} by directly applying the Yangian level-one generators to the Grassmannian integral itself.


In \cite{Drummond:2010qh} it was found that applying the level-one generator to the form $K$ yields a total derivative,
\begin{equation}
J^{(1)}{}^{\AA}{}_{\BB} K = d \Omega^{\AA}{}_{\BB}.
\label{exactvar}
\end{equation}
This property guarantees that $\mathcal{L}$ is invariant for every choice of closed contour. Moreover it has been shown \cite{Drummond:2010uq,Korchemsky:2010ut} that the form $K$ is unique after imposing the condition (\ref{exactvar}). In this sense the Grassmannian integral is the most general form of Yangian invariant. Moreover, replacing $\delta^{4|4} \longrightarrow \delta^{m|m}$, the formulas (\ref{ACCK},\ref{MS}) are equally valid for generating invariants of the symmetry $Y(psl(m|m))$ where one no longer has the interpretation of the symmetry as superconformal symmetry. Thus the Grassmannian integral formula is really naturally associated to the series of Yangians $Y(psl(m|m))$.

It is very striking that the leading singularities seem to be all given by Yangian invariants and even more striking that they seem to exhaust all such possibilities. 
The first of these statements follows from the analysis of leading singularities in \cite{ArkaniHamed:2010kv,Drummond:2010mb}. The second still requires rigorous proof but is consistent with all investigations so far conducted of leading singularities and residues in the Grassmannian. In some sense one can say that the leading singularity part of the amplitude is being determined by its symmetry. 
In fact the invariance of the leading singularities follows from the fact that the all-loop planar integrand is Yangian invariant up to a total derivative. This was shown by constructing it via a BCFW type recursion relation in a way which respects the Yangian symmetry \cite{ArkaniHamed:2010kv}.

It is not yet clear if the full Yangian invariance $Y(psl(4|4))$ exhibits itself on the actual amplitudes themselves (i.e. after the loop integrations have been performed). As we have discussed the problem lies in the breakdown of the original (super)conformal symmetry.
At one loop for MHV amplitudes (or Wilson loops) a $Y(sl(2)) \oplus Y(sl(2))$ subalgebra of the full symmetry is present \cite{Drummond:2010zv} for a special restricted two-dimensional kinematical setup \cite{Alday:2009yn}. It is possible to consider a particular finite, conformally invariant ratio of light-like Wilson loops, introduced in \cite{Alday:2010ku,Gaiotto:2010fk} in order to understand the OPE properties of light-like Wilson loops. The conformal symmetry of this setup is $sl(2)\oplus sl(2)$ and on the finite ratio the two commuting symmetries each extend to their Yangians in a natural way. This is equivalent to the effects of the original conformal symmetry in the two-dimensional kinematics.
It remains to be seen to what extent this statement can be extended beyond the restricted kinematics and beyond one loop. Since the symmetry manifests itself as certain second-order differential equations it is possible that the differential equations found in \cite{Drummond:2010cz} for certain momentum twistor loop integrals will be important in understanding whether this can be implemented.
If the Yangian structure does manifest itself on all the loop corrections this would in some sense amount to the integrability of the S-matrix of planar $\mathcal{N}=4$ super Yang-Mills theory.

\section*{Acknowledgements}
It is a pleasure to thank Livia Ferro, Johannes Henn, Gregory Korchemsky, Jan Plefka, Vladimir Smirnov and Emery Sokatchev for collaboration on the topics presented in this review.

\phantomsection
\addcontentsline{toc}{section}{\refname}
\bibliography{final}

\begin{thebibliography}{VI.1}
\ifx\href\asklfhas\newcommand{\href}[2]{#2}\fi
\ifx\arxivref\asklfhas\newcommand{\arxivref}[2]{\href{http://arxiv.org/abs/#1}{#2}}\fi
\ifx\doiref\asklfhas\newcommand{\doiref}[2]{\href{http://dx.doi.org/#1}{#2}}\fi
\raggedright
\small
\parskip 0pt

\bibitem[V.1]{chapAmp}
R.~Roiban,
\textit{``Review of AdS/CFT Integrability, Chapter V.1: Scattering Amplitudes
  -- Formalism, Recursion and Unitarity''}.

\bibitem[V.3]{chapTdual}
L.~F.~Alday,
\textit{``Review of AdS/CFT Integrability, Chapter V.3: Scattering Amplitudes
  at Strong Coupling''}.

\bibitem[VI.1]{chapSuperconf}
N.~Beisert,
\textit{``Review of AdS/CFT Integrability, Chapter VI.1: Superconformal
  Algebra''}.

\bibitem{Anastasiou:2003kj}
C.~Anastasiou, Z.~Bern, L.~J.~Dixon and D.~A.~Kosower,
\textit{``{Planar amplitudes in maximally supersymmetric Yang-Mills theory}''},
\textsf{\doiref{10.1103/PhysRevLett.91.251602}{Phys.~Rev.~Lett.~91,~251602~(2003)}},
\texttt{\arxivref{hep-th/0309040}{hep-th/0309040}}.

\bibitem{Bern:2010qa}
Z.~Bern, J.~J.~Carrasco, T.~Dennen, Y.-t.~Huang and H.~Ita,
\textit{``{Generalized Unitarity and Six-Dimensional Helicity}''},
\texttt{\arxivref{1010.0494}{arxiv:1010.0494}}.

\bibitem{CaronHuot:2010rj}
S.~Caron-Huot and D.~O'Connell,
\textit{``{Spinor Helicity and Dual Conformal Symmetry in Ten Dimensions}''},
\texttt{\arxivref{1010.5487}{arxiv:1010.5487}}.

\bibitem{Dennen:2010dh}
T.~Dennen and Y.-t.~Huang,
\textit{``{Dual Conformal Properties of Six-Dimensional Maximal Super
  Yang-Mills Amplitudes}''},
\texttt{\arxivref{1010.5874}{arxiv:1010.5874}}.

\bibitem{Akhoury:1978vq}
R.~Akhoury,
\textit{``{MASS DIVERGENCES OF WIDE ANGLE SCATTERING AMPLITUDES}''},
\textsf{\doiref{10.1103/PhysRevD.19.1250}{Phys.~Rev.~D19,~1250~(1979)}}.

\bibitem{Mueller:1979ih}
A.~H.~Mueller,
\textit{``{ON THE ASYMPTOTIC BEHAVIOR OF THE SUDAKOV FORM-FACTOR}''},
\textsf{\doiref{10.1103/PhysRevD.20.2037}{Phys.~Rev.~D20,~2037~(1979)}}.

\bibitem{Collins:1980ih}
J.~C.~Collins,
\textit{``{ALGORITHM TO COMPUTE CORRECTIONS TO THE SUDAKOV FORM- FACTOR}''},
\textsf{\doiref{10.1103/PhysRevD.22.1478}{Phys.~Rev.~D22,~1478~(1980)}}.

\bibitem{Sen:1981sd}
A.~Sen,
\textit{``{Asymptotic Behavior of the Sudakov Form-Factor in QCD}''},
\textsf{\doiref{10.1103/PhysRevD.24.3281}{Phys.~Rev.~D24,~3281~(1981)}}.

\bibitem{Sterman:1986aj}
G.~Sterman,
\textit{``{Summation of Large Corrections to Short Distance Hadronic
  Cross-Sections}''},
\textsf{\doiref{10.1016/0550-3213(87)90258-6}{Nucl.~Phys.~B281,~310~(1987)}}.

\bibitem{Botts:1989kf}
J.~Botts and G.~Sterman,
\textit{``{Hard Elastic Scattering in QCD: Leading Behavior}''},
\textsf{\doiref{10.1016/0550-3213(89)90372-6}{Nucl.~Phys.~B325,~62~(1989)}}.

\bibitem{Catani:1989ne}
S.~Catani and L.~Trentadue,
\textit{``{Resummation of the QCD Perturbative Series for Hard Processes}''},
\textsf{\doiref{10.1016/0550-3213(89)90273-3}{Nucl.~Phys.~B327,~323~(1989)}}.

\bibitem{Bern:2005iz}
Z.~Bern, L.~J.~Dixon and V.~A.~Smirnov,
\textit{``{Iteration of planar amplitudes in maximally supersymmetric
  Yang-Mills theory at three loops and beyond}''},
\textsf{\doiref{10.1103/PhysRevD.72.085001}{Phys.~Rev.~D72,~085001~(2005)}},
\texttt{\arxivref{hep-th/0505205}{hep-th/0505205}}.

\bibitem{Korchemsky:1988hd}
G.~P.~Korchemsky,
\textit{``{SUDAKOV FORM-FACTOR IN QCD}''},
\textsf{\doiref{10.1016/0370-2693(89)90799-5}{Phys.~Lett.~B220,~629~(1989)}}.

\bibitem{Korchemsky:1988pn}
G.~P.~Korchemsky,
\textit{``{DOUBLE LOGARITHMIC ASYMPTOTICS IN QCD}''},
\textsf{\doiref{10.1016/0370-2693(89)90876-9}{Phys.~Lett.~B217,~330~(1989)}}.

\bibitem{Magnea:1990zb}
L.~Magnea and G.~Sterman,
\textit{``{Analytic continuation of the Sudakov form-factor in QCD}''},
\textsf{\doiref{10.1103/PhysRevD.42.4222}{Phys.~Rev.~D42,~4222~(1990)}}.

\bibitem{Korchemsky:1993uz}
G.~P.~Korchemsky and G.~Marchesini,
\textit{``{Resummation of large infrared corrections using Wilson loops}''},
\textsf{\doiref{10.1016/0370-2693(93)90015-A}{Phys.~Lett.~B313,~433~(1993)}}.

\bibitem{Catani:1998bh}
S.~Catani,
\textit{``{The singular behaviour of {QCD} amplitudes at two-loop order}''},
\textsf{\doiref{10.1016/S0370-2693(98)00332-3}{Phys.~Lett.~B427,~161~(1998)}},
\texttt{\arxivref{hep-ph/9802439}{hep-ph/9802439}}.

\bibitem{Sterman:2002qn}
G.~Sterman and M.~E.~Tejeda-Yeomans,
\textit{``{Multi-loop amplitudes and resummation}''},
\textsf{\doiref{10.1016/S0370-2693(02)03100-3}{Phys.~Lett.~B552,~48~(2003)}},
\texttt{\arxivref{hep-ph/0210130}{hep-ph/0210130}}.

\bibitem{Ivanov:1985np}
S.~V.~Ivanov, G.~P.~Korchemsky and A.~V.~Radyushkin,
\textit{``{INFRARED ASYMPTOTICS OF PERTURBATIVE QCD: CONTOUR GAUGES}''},
\textsf{Yad.~Fiz.~44,~230~(1986)}.

\bibitem{Korchemsky:1992xv}
G.~P.~Korchemsky and G.~Marchesini,
\textit{``{Structure function for large x and renormalization of Wilson
  loop}''},
\textsf{\doiref{10.1016/0550-3213(93)90167-N}{Nucl.~Phys.~B406,~225~(1993)}},
\texttt{\arxivref{hep-ph/9210281}{hep-ph/9210281}}.

\bibitem{Cachazo:2006tj}
F.~Cachazo, M.~Spradlin and A.~Volovich,
\textit{``{Iterative structure within the five-particle two-loop amplitude}''},
\textsf{\doiref{10.1103/PhysRevD.74.045020}{Phys.~Rev.~D74,~045020~(2006)}},
\texttt{\arxivref{hep-th/0602228}{hep-th/0602228}}.

\bibitem{Bern:2006vw}
Z.~Bern, M.~Czakon, D.~A.~Kosower, R.~Roiban and V.~A.~Smirnov,
\textit{``{Two-loop iteration of five-point $\mathcal{N}$ = 4 super-Yang-Mills
  amplitudes}''},
\textsf{\doiref{10.1103/PhysRevLett.97.181601}{Phys.~Rev.~Lett.~97,~181601~(2006)}},
\texttt{\arxivref{hep-th/0604074}{hep-th/0604074}}.

\bibitem{Drummond:2006rz}
J.~M.~Drummond, J.~Henn, V.~A.~Smirnov and E.~Sokatchev,
\textit{``{Magic identities for conformal four-point integrals}''},
\textsf{\doiref{10.1088/1126-6708/2007/01/064}{JHEP~0701,~064~(2007)}},
\texttt{\arxivref{hep-th/0607160}{hep-th/0607160}}.

\bibitem{Spradlin:2008uu}
M.~Spradlin, A.~Volovich and C.~Wen,
\textit{``{Three-Loop Leading Singularities and BDS Ansatz for Five
  Particles}''},
\textsf{\doiref{10.1103/PhysRevD.78.085025}{Phys.~Rev.~D78,~085025~(2008)}},
\texttt{\arxivref{0808.1054}{arxiv:0808.1054}}.

\bibitem{Korchemskaya:1996je}
I.~A.~Korchemskaya and G.~P.~Korchemsky,
\textit{``{Evolution equation for gluon Regge trajectory}''},
\textsf{\doiref{10.1016/0370-2693(96)01016-7}{Phys.~Lett.~B387,~346~(1996)}},
\texttt{\arxivref{hep-ph/9607229}{hep-ph/9607229}}.

\bibitem{Alday:2007hr}
L.~F.~Alday and J.~M.~Maldacena,
\textit{``{Gluon scattering amplitudes at strong coupling}''},
\textsf{\doiref{10.1088/1126-6708/2007/06/064}{JHEP~0706,~064~(2007)}},
\texttt{\arxivref{0705.0303}{arxiv:0705.0303}}.

\bibitem{Drummond:2007aua}
J.~M.~Drummond, G.~P.~Korchemsky and E.~Sokatchev,
\textit{``{Conformal properties of four-gluon planar amplitudes and Wilson
  loops}''},
\textsf{\doiref{10.1016/j.nuclphysb.2007.11.041}{Nucl.~Phys.~B795,~385~(2008)}},
\texttt{\arxivref{0707.0243}{arxiv:0707.0243}}.

\bibitem{Brandhuber:2007yx}
A.~Brandhuber, P.~Heslop and G.~Travaglini,
\textit{``{MHV Amplitudes in $\mathcal{N}$ = 4 Super Yang-Mills and Wilson
  Loops}''},
\textsf{\doiref{10.1016/j.nuclphysb.2007.11.002}{Nucl.~Phys.~B794,~231~(2008)}},
\texttt{\arxivref{0707.1153}{arxiv:0707.1153}}.

\bibitem{Drummond:2007cf}
J.~M.~Drummond, J.~Henn, G.~P.~Korchemsky and E.~Sokatchev,
\textit{``{On planar gluon amplitudes/Wilson loops duality}''},
\textsf{\doiref{10.1016/j.nuclphysb.2007.11.007}{Nucl.~Phys.~B795,~52~(2008)}},
\texttt{\arxivref{0709.2368}{arxiv:0709.2368}}.

\bibitem{Drummond:2007au}
J.~M.~Drummond, J.~Henn, G.~P.~Korchemsky and E.~Sokatchev,
\textit{``{Conformal Ward identities for Wilson loops and a test of the duality
  with gluon amplitudes}''},
\texttt{\arxivref{0712.1223}{arxiv:0712.1223}}.

\bibitem{Drummond:2007bm}
J.~M.~Drummond, J.~Henn, G.~P.~Korchemsky and E.~Sokatchev,
\textit{``{The hexagon Wilson loop and the BDS ansatz for the six-gluon
  amplitude}''},
\textsf{\doiref{10.1016/j.physletb.2008.03.032}{Phys.~Lett.~B662,~456~(2008)}},
\texttt{\arxivref{0712.4138}{arxiv:0712.4138}}.

\bibitem{Drummond:2008aq}
J.~M.~Drummond, J.~Henn, G.~P.~Korchemsky and E.~Sokatchev,
\textit{``{Hexagon Wilson loop = six-gluon MHV amplitude}''},
\textsf{\doiref{10.1016/j.nuclphysb.2009.02.015}{Nucl.~Phys.~B815,~142~(2009)}},
\texttt{\arxivref{0803.1466}{arxiv:0803.1466}}.

\bibitem{Bern:2008ap}
Z.~Bern, L.~Dixon, D.~Kosower, R.~Roiban, M.~Spradlin, C.~Vergu and
  A.~Volovich,
\textit{``{The Two-Loop Six-Gluon MHV Amplitude in Maximally Supersymmetric
  Yang-Mills Theory}''},
\textsf{\doiref{10.1103/PhysRevD.78.045007}{Phys.~Rev.~D78,~045007~(2008)}},
\texttt{\arxivref{0803.1465}{arxiv:0803.1465}}.

\bibitem{Bern:2006ew}
Z.~Bern, M.~Czakon, L.~J.~Dixon, D.~A.~Kosower and V.~A.~Smirnov,
\textit{``{The Four-Loop Planar Amplitude and Cusp Anomalous Dimension in
  Maximally Supersymmetric Yang-Mills Theory}''},
\textsf{\doiref{10.1103/PhysRevD.75.085010}{Phys.~Rev.~D75,~085010~(2007)}},
\texttt{\arxivref{hep-th/0610248}{hep-th/0610248}}.

\bibitem{DelDuca:2010zg}
V.~Del~Duca, C.~Duhr and V.~A.~Smirnov,
\textit{``{The Two-Loop Hexagon Wilson Loop in N = 4 SYM}''},
\texttt{\arxivref{1003.1702}{arxiv:1003.1702}}.

\bibitem{Anastasiou:2009kna}
C.~Anastasiou, A.~Brandhuber, P.~Heslop, V.~V.~Khoze, B.~Spence and
  G.~Travaglini,
\textit{``{Two-Loop Polygon Wilson Loops in $\mathcal{N}$ = 4 SYM}''},
\textsf{\doiref{10.1088/1126-6708/2009/05/115}{JHEP~0905,~115~(2009)}},
\texttt{\arxivref{0902.2245}{arxiv:0902.2245}}.

\bibitem{Vergu:2009zm}
C.~Vergu,
\textit{``{Higher point MHV amplitudes in $\mathcal{N}$ = 4 Supersymmetric
  Yang-Mills Theory}''},
\textsf{\doiref{10.1103/PhysRevD.79.125005}{Phys.~Rev.~D79,~125005~(2009)}},
\texttt{\arxivref{0903.3526}{arxiv:0903.3526}}.

\bibitem{Vergu:2009tu}
C.~Vergu,
\textit{``{The two-loop MHV amplitudes in $\mathcal{N}$ = 4 supersymmetric
  Yang-Mills theory}''},
\texttt{\arxivref{0908.2394}{arxiv:0908.2394}}.

\bibitem{Mason:2010yk}
L.~Mason and D.~Skinner,
\textit{``{The Complete Planar S-matrix of N=4 SYM as a Wilson Loop in Twistor
  Space}''},
\textsf{\doiref{10.1007/JHEP12(2010)018}{JHEP~1012,~018~(2010)}},
\texttt{\arxivref{1009.2225}{arxiv:1009.2225}}.

\bibitem{CaronHuot:2010ek}
S.~Caron-Huot,
\textit{``{Notes on the scattering amplitude / Wilson loop duality}''},
\texttt{\arxivref{1010.1167}{arxiv:1010.1167}}.

\bibitem{Drummond:2008cr}
J.~M.~Drummond and J.~M.~Henn,
\textit{``{All tree-level amplitudes in $\mathcal{N}$ = 4 SYM}''},
\textsf{\doiref{10.1088/1126-6708/2009/04/018}{JHEP~0904,~018~(2009)}},
\texttt{\arxivref{0808.2475}{arxiv:0808.2475}}.

\bibitem{Brandhuber:2008pf}
A.~Brandhuber, P.~Heslop and G.~Travaglini,
\textit{``{A note on dual superconformal symmetry of the $\mathcal{N}$ = 4
  super Yang-Mills S-matrix}''},
\textsf{\doiref{10.1103/PhysRevD.78.125005}{Phys.~Rev.~D78,~125005~(2008)}},
\texttt{\arxivref{0807.4097}{arxiv:0807.4097}}.

\bibitem{ArkaniHamed:2008gz}
N.~Arkani-Hamed, F.~Cachazo and J.~Kaplan,
\textit{``{What is the Simplest Quantum Field Theory?}''},
\texttt{\arxivref{0808.1446}{arxiv:0808.1446}}.

\bibitem{Elvang:2008na}
H.~Elvang, D.~Z.~Freedman and M.~Kiermaier,
\textit{``{Recursion Relations, Generating Functions, and Unitarity Sums in N=4
  SYM Theory}''},
\textsf{\doiref{10.1088/1126-6708/2009/04/009}{JHEP~0904,~009~(2009)}},
\texttt{\arxivref{0808.1720}{arxiv:0808.1720}}.

\bibitem{Bern:2007ct}
Z.~Bern, J.~J.~M.~Carrasco, H.~Johansson and D.~A.~Kosower,
\textit{``{Maximally supersymmetric planar Yang-Mills amplitudes at five
  loops}''},
\textsf{\doiref{10.1103/PhysRevD.76.125020}{Phys.~Rev.~D76,~125020~(2007)}},
\texttt{\arxivref{0705.1864}{arxiv:0705.1864}}.

\bibitem{Britto:2004ap}
R.~Britto, F.~Cachazo and B.~Feng,
\textit{``{New Recursion Relations for Tree Amplitudes of Gluons}''},
\textsf{\doiref{10.1016/j.nuclphysb.2005.02.030}{Nucl.~Phys.~B715,~499~(2005)}},
\texttt{\arxivref{hep-th/0412308}{hep-th/0412308}}.

\bibitem{Britto:2005fq}
R.~Britto, F.~Cachazo, B.~Feng and E.~Witten,
\textit{``{Direct Proof Of Tree-Level Recursion Relation In Yang- Mills
  Theory}''},
\textsf{\doiref{10.1103/PhysRevLett.94.181602}{Phys.~Rev.~Lett.~94,~181602~(2005)}},
\texttt{\arxivref{hep-th/0501052}{hep-th/0501052}}.

\bibitem{Witten:2003nn}
E.~Witten,
\textit{``{Perturbative gauge theory as a string theory in twistor space}''},
\textsf{\doiref{10.1007/s00220-004-1187-3}{Commun.~Math.~Phys.~252,~189~(2004)}},
\texttt{\arxivref{hep-th/0312171}{hep-th/0312171}}.

\bibitem{Bargheer:2009qu}
T.~Bargheer, N.~Beisert, W.~Galleas, F.~Loebbert and T.~McLoughlin,
\textit{``{Exacting $\mathcal{N}$ = 4 Superconformal Symmetry}''},
\texttt{\arxivref{0905.3738}{arxiv:0905.3738}}.

\bibitem{Korchemsky:2009hm}
G.~P.~Korchemsky and E.~Sokatchev,
\textit{``{Symmetries and analytic properties of scattering amplitudes in
  $\mathcal{N}$ = 4 SYM theory}''},
\texttt{\arxivref{0906.1737}{arxiv:0906.1737}}.

\bibitem{Sever:2009aa}
A.~Sever and P.~Vieira,
\textit{``{Symmetries of the $\mathcal{N}$ = 4 SYM S-matrix}''},
\texttt{\arxivref{0908.2437}{arxiv:0908.2437}}.

\bibitem{Drummond:2008vq}
J.~M.~Drummond, J.~Henn, G.~P.~Korchemsky and E.~Sokatchev,
\textit{``{Dual superconformal symmetry of scattering amplitudes in
  $\mathcal{N}$ = 4 super-Yang-Mills theory}''},
\texttt{\arxivref{0807.1095}{arxiv:0807.1095}}.

\bibitem{Drummond:2008bq}
J.~M.~Drummond, J.~Henn, G.~P.~Korchemsky and E.~Sokatchev,
\textit{``{Generalized unitarity for $\mathcal{N}$ = 4 super-amplitudes}''},
\texttt{\arxivref{0808.0491}{arxiv:0808.0491}}.

\bibitem{Bern:2004bt}
Z.~Bern, L.~J.~Dixon and D.~A.~Kosower,
\textit{``{All next-to-maximally helicity-violating one-loop gluon amplitudes
  in $\mathcal{N}$ = 4 super-Yang-Mills theory}''},
\textsf{\doiref{10.1103/PhysRevD.72.045014}{Phys.~Rev.~D72,~045014~(2005)}},
\texttt{\arxivref{hep-th/0412210}{hep-th/0412210}}.

\bibitem{Brandhuber:2009kh}
A.~Brandhuber, P.~Heslop and G.~Travaglini,
\textit{``{Proof of the Dual Conformal Anomaly of One-Loop Amplitudes in
  $\mathcal{N}$ = 4 SYM}''},
\texttt{\arxivref{0906.3552}{arxiv:0906.3552}}.

\bibitem{Alday:2009zm}
L.~F.~Alday, J.~M.~Henn, J.~Plefka and T.~Schuster,
\textit{``{Scattering into the fifth dimension of $\mathcal{N}$ = 4 super
  Yang-Mills}''},
\texttt{\arxivref{0908.0684}{arxiv:0908.0684}}.

\bibitem{Beisert:2010gn}
N.~Beisert, J.~Henn, T.~McLoughlin and J.~Plefka,
\textit{``{One-Loop Superconformal and Yangian Symmetries of Scattering
  Amplitudes in N=4 Super Yang-Mills}''},
\textsf{\doiref{10.1007/JHEP04(2010)085}{JHEP~1004,~085~(2010)}},
\texttt{\arxivref{1002.1733}{arxiv:1002.1733}}.

\bibitem{Drummond:2009fd}
J.~M.~Drummond, J.~M.~Henn and J.~Plefka,
\textit{``{Yangian symmetry of scattering amplitudes in $\mathcal{N}$ = 4 super
  Yang-Mills theory}''},
\textsf{\doiref{10.1088/1126-6708/2009/05/046}{JHEP~0905,~046~(2009)}},
\texttt{\arxivref{0902.2987}{arxiv:0902.2987}}.

\bibitem{Dolan:2003uh}
L.~Dolan, C.~R.~Nappi and E.~Witten,
\textit{``{A relation between approaches to integrability in superconformal
  Yang-Mills theory}''},
\textsf{\doiref{10.1088/1126-6708/2003/10/017}{JHEP~0310,~017~(2003)}},
\texttt{\arxivref{hep-th/0308089}{hep-th/0308089}}.

\bibitem{Dolan:2004ps}
L.~Dolan, C.~R.~Nappi and E.~Witten,
\textit{``{Yangian symmetry in D=4 superconformal Yang-Mills theory}''},
\texttt{\arxivref{hep-th/0401243}{hep-th/0401243}}.

\bibitem{Drummond:2010qh}
J.~M.~Drummond and L.~Ferro,
\textit{``{Yangians, Grassmannians and T-duality}''},
\texttt{\arxivref{1001.3348}{arxiv:1001.3348}}.

\bibitem{Hodges:2009hk}
A.~Hodges,
\textit{``{Eliminating spurious poles from gauge-theoretic amplitudes}''},
\texttt{\arxivref{0905.1473}{arxiv:0905.1473}}.

\bibitem{Berkovits:2008ic}
N.~Berkovits and J.~Maldacena,
\textit{``{Fermionic T-Duality, Dual Superconformal Symmetry, and the
  Amplitude/Wilson Loop Connection}''},
\textsf{\doiref{10.1088/1126-6708/2008/09/062}{JHEP~0809,~062~(2008)}},
\texttt{\arxivref{0807.3196}{arxiv:0807.3196}}.

\bibitem{Beisert:2008iq}
N.~Beisert, R.~Ricci, A.~A.~Tseytlin and M.~Wolf,
\textit{``{Dual Superconformal Symmetry from AdS$_5$ $\times$ S$^5$ Superstring
  Integrability}''},
\textsf{\doiref{10.1103/PhysRevD.78.126004}{Phys.~Rev.~D78,~126004~(2008)}},
\texttt{\arxivref{0807.3228}{arxiv:0807.3228}}.

\bibitem{Beisert:2009cs}
N.~Beisert,
\textit{``{T-Duality, Dual Conformal Symmetry and Integrability for Strings on
  AdS$_5$ $\times$ S$^5$}''},
\textsf{Fortschr.~Phys.~57,~329~(2009)},
\texttt{\arxivref{0903.0609}{arxiv:0903.0609}}.

\bibitem{Korchemsky:2009jv}
G.~P.~Korchemsky and E.~Sokatchev,
\textit{``{Twistor transform of all tree amplitudes in $\mathcal{N}$ = 4 SYM
  theory}''},
\texttt{\arxivref{0907.4107}{arxiv:0907.4107}}.

\bibitem{Henn:2010bk}
J.~M.~Henn, S.~G.~Naculich, H.~J.~Schnitzer and M.~Spradlin,
\textit{``{Higgs-regularized three-loop four-gluon amplitude in N=4 SYM:
  exponentiation and Regge limits}''},
\textsf{\doiref{10.1007/JHEP04(2010)038}{JHEP~1004,~038~(2010)}},
\texttt{\arxivref{1001.1358}{arxiv:1001.1358}}.

\bibitem{Drummond:2010uq}
J.~M.~Drummond and L.~Ferro,
\textit{``{The Yangian origin of the Grassmannian integral}''},
\texttt{\arxivref{1002.4622}{arxiv:1002.4622}}.

\bibitem{Korchemsky:2010ut}
G.~P.~Korchemsky and E.~Sokatchev,
\textit{``{Superconformal invariants for scattering amplitudes in N=4 SYM
  theory}''},
\texttt{\arxivref{1002.4625}{arxiv:1002.4625}}.

\bibitem{ArkaniHamed:2009dn}
N.~Arkani-Hamed, F.~Cachazo, C.~Cheung and J.~Kaplan,
\textit{``{A Duality For The S Matrix}''},
\texttt{\arxivref{0907.5418}{arxiv:0907.5418}}.

\bibitem{Mason:2009qx}
L.~Mason and D.~Skinner,
\textit{``{Dual Superconformal Invariance, Momentum Twistors and
  Grassmannians}''},
\texttt{\arxivref{0909.0250}{arxiv:0909.0250}}.

\bibitem{ArkaniHamed:2009vw}
N.~Arkani-Hamed, F.~Cachazo and C.~Cheung,
\textit{``{The Grassmannian Origin Of Dual Superconformal Invariance}''},
\textsf{\doiref{10.1007/JHEP03(2010)036}{JHEP~1003,~036~(2010)}},
\texttt{\arxivref{0909.0483}{arxiv:0909.0483}}.

\bibitem{ArkaniHamed:2010kv}
N.~Arkani-Hamed, J.~L.~Bourjaily, F.~Cachazo, S.~Caron-Huot and J.~Trnka,
\textit{``{The All-Loop Integrand For Scattering Amplitudes in Planar N=4
  SYM}''},
\texttt{\arxivref{1008.2958}{arxiv:1008.2958}}.

\bibitem{Drummond:2010mb}
J.~M.~Drummond and J.~M.~Henn,
\textit{``{Simple loop integrals and amplitudes in N=4 SYM}''},
\texttt{\arxivref{1008.2965}{arxiv:1008.2965}}.

\bibitem{Drummond:2010zv}
J.~M.~Drummond, L.~Ferro and E.~Ragoucy,
\textit{``{Yangian symmetry of light-like Wilson loops}''},
\texttt{\arxivref{1011.4264}{arxiv:1011.4264}}.

\bibitem{Alday:2009yn}
L.~F.~Alday and J.~Maldacena,
\textit{``{Null polygonal Wilson loops and minimal surfaces in Anti-de-Sitter
  space}''},
\texttt{\arxivref{0904.0663}{arxiv:0904.0663}}.

\bibitem{Alday:2010ku}
L.~F.~Alday, D.~Gaiotto, J.~Maldacena, A.~Sever and P.~Vieira,
\textit{``{An Operator Product Expansion for Polygonal null Wilson Loops}''},
\texttt{\arxivref{1006.2788}{arxiv:1006.2788}}.

\bibitem{Boels:2010mj}
R.~H.~Boels,
\textit{``{No triangles on the moduli space of maximally supersymmetric gauge
  theory}''},
\textsf{\doiref{10.1007/JHEP05(2010)046}{JHEP~1005,~046~(2010)}},
\texttt{\arxivref{1003.2989}{arxiv:1003.2989}}.

\bibitem{Gaiotto:2010fk}
D.~Gaiotto, J.~Maldacena, A.~Sever and P.~Vieira,
\textit{``{Bootstrapping Null Polygon Wilson Loops}''},
\texttt{\arxivref{1010.5009}{arxiv:1010.5009}}.

\bibitem{Drummond:2010cz}
J.~M.~Drummond, J.~M.~Henn and J.~Trnka,
\textit{``{New differential equations for on-shell loop integrals}''},
\texttt{\arxivref{1010.3679}{arxiv:1010.3679}}.

\bibitem{Henn:2010ir}
J.~M.~Henn, S.~G.~Naculich, H.~J.~Schnitzer and M.~Spradlin,
\textit{``{More loops and legs in Higgs-regulated N=4 SYM amplitudes}''},
\textsf{\doiref{10.1007/JHEP08(2010)002}{JHEP~1008,~002~(2010)}},
\texttt{\arxivref{1004.5381}{arxiv:1004.5381}}.

\end{thebibliography}
\bibliographystyle{nbc}

\end{document}